\documentclass[sigconf]{acmart}

\AtBeginDocument{%
  }

\copyrightyear{2026}
\acmYear{2026}
\setcopyright{cc}
\setcctype{by}
\acmConference[SIGMOD Companion '26]{Companion of the International Conference on Management of Data}{May 31--June 5, 2026}{Bengaluru, India.}
\acmBooktitle{Companion of the International Conference on Management of Data (SIGMOD Companion '26), May 31--June 5, 2026, Bengaluru, India}
\acmDOI{10.1145/3788853.3803093}
\acmISBN{979-8-4007-2450-3/2026/05}
\usepackage{algorithm}
\usepackage{algpseudocode}
\usepackage{comment}
\usepackage{multirow}
\usepackage{booktabs}
\usepackage{amsmath}

\usepackage{amssymb}
\usepackage{float} 
\usepackage{graphicx}
\graphicspath{ {./} }
\usepackage{listings}
\usepackage{xcolor}

\usepackage{multicol}
\usepackage{needspace}
\usepackage{balance}

\definecolor{codegray}{rgb}{0.5,0.5,0.5}
\definecolor{codepurple}{rgb}{0.58,0,0.82}
\definecolor{backcolour}{rgb}{0.95,0.95,0.92}

\lstdefinelanguage{SQL}{
  morekeywords={
    SELECT,FROM,WHERE,GROUP,BY,ORDER,HAVING,AS,ON,JOIN,INNER,LEFT,RIGHT,
    OUTER,UNION,ALL,INSERT,INTO,VALUES,UPDATE,SET,DELETE,CREATE,DROP,
    TABLE,VIEW,INDEX,AND,OR,NOT,NULL,IS,DISTINCT,CASE,WHEN,THEN,END,
    COUNT,
    PROMPT,FILE,TO_FILE,
    AI\_FILTER,AI\_AGG,AI\_CLASSIFY,AI\_COMPLETE,AI\_SUMMARIZE\_AGG
  },
  sensitive=false,
  morecomment=[l]{--},
  morestring=[b]',
}

\lstdefinestyle{sqlstyle}{
  commentstyle=\color{gray}\itshape,
  keywordstyle=\color{blue}\bfseries,
  stringstyle=\color{codepurple},
  basicstyle=\ttfamily\small,
  breaklines=true,
  frame=single,
  showstringspaces=false,
  captionpos=b,
  columns=flexible,
}

\begin{document}

\title{Cortex AISQL: A Production SQL Engine for Unstructured Data}

\author{Paweł Liskowski}
\email{pawel.liskowski@snowflake.com}
\affiliation{%
  \institution{Snowflake Inc.}
  \city{Poznań}
  \country{Poland}
}

\author{Benjamin Han}
\email{benjamin.han@snowflake.com}
\affiliation{%
  \institution{Snowflake Inc.}
  \city{Menlo Park}
  \state{CA}
  \country{USA}
}

\author{Paritosh Aggarwal}
\email{paritosh.aggarwal@snowflake.com}
\affiliation{%
  \institution{Snowflake Inc.}
  \city{Menlo Park}
  \state{CA}
  \country{USA}
}

\author{Bowei Chen}
\email{bowei.chen@snowflake.com}
\affiliation{%
  \institution{Snowflake Inc.}
  \city{Menlo Park}
  \state{CA}
  \country{USA}
}

\author{Boxin Jiang}
\email{boxin.jiang@snowflake.com}
\affiliation{%
  \institution{Snowflake Inc.}
  \city{Menlo Park}
  \state{CA}
  \country{USA}
}

\author{Nitish Jindal}
\email{nitish.jindal@snowflake.com}
\affiliation{%
  \institution{Snowflake Inc.}
  \city{Redmond}
  \state{WA}
  \country{USA}
}

\author{Zihan Li}
\email{zihan.li@snowflake.com}
\affiliation{%
  \institution{Snowflake Inc.}
  \city{Menlo Park}
  \state{CA}
  \country{USA}
}

\author{Aaron Lin}
\email{aaron.lin@snowflake.com}
\affiliation{%
  \institution{Snowflake Inc.}
  \city{Menlo Park}
  \state{CA}
  \country{USA}
}

\author{Kyle Schmaus}
\email{kyle.schmaus@snowflake.com}
\affiliation{%
  \institution{Snowflake Inc.}
  \city{Menlo Park}
  \state{CA}
  \country{USA}
}

\author{Jay Tayade}
\email{jay.tayade@snowflake.com}
\affiliation{%
  \institution{Snowflake Inc.}
  \city{Seattle}
  \state{WA}
  \country{USA}
}

\author{Weicheng Zhao}
\email{weicheng.zhao@snowflake.com}
\affiliation{%
  \institution{Snowflake Inc.}
  \city{Menlo Park}
  \state{CA}
  \country{USA}
}

\author{Anupam Datta}
\email{anupam.datta@snowflake.com}
\affiliation{%
  \institution{Snowflake Inc.}
  \city{Menlo Park}
  \state{CA}
  \country{USA}
}

\author{Nathan Wiegand}
\email{nathan.wiegand@snowflake.com}
\affiliation{%
  \institution{Snowflake Inc.}
  \city{Austin}
  \state{TX}
  \country{USA}
}

\author{Dimitris Tsirogiannis}
\email{dimitris.tsirogiannis@snowflake.com}
\affiliation{%
  \institution{Snowflake Inc.}
  \city{Menlo Park}
  \state{CA}
  \country{USA}
}

\renewcommand{\shortauthors}{Paweł Liskowski et al.}

\begin{abstract}
Integrating AI's semantic capabilities directly into SQL is a key goal, as it would enable users to write declarative queries that blend relational operations with semantic reasoning over both structured and unstructured data. However, achieving production-scale efficiency for semantic operations presents significant hurdles. Semantic operations inherently cost more than traditional SQL operations and possess different latency and throughput characteristics, making their application at large scales prohibitively expensive. Current query execution engines are not designed to optimize semantic operations, often leading to suboptimal query plans. The challenge is compounded because the cost and selectivity of semantic operations are typically unknown at compilation time. Snowflake's Cortex AISQL is a query execution engine that addresses these challenges through three novel techniques informed by production deployment data from Snowflake customers. First, AI-aware query optimization treats AI inference cost as a first-class optimization objective, reasoning about large language model (LLM) cost directly during query planning to achieve 2--8$\times$ speedups. Second, adaptive model cascades reduce inference costs by routing most rows through a fast proxy model while escalating uncertain cases to a powerful oracle model, with 2--6$\times$ speedups at 90--95\% of oracle model quality. Third, semantic join query rewriting lowers the quadratic time complexity of join operations to linear through reformulation as multi-label classification tasks, for 15--70$\times$ speedups, often with improved prediction quality. AISQL is deployed in production at Snowflake, where it powers diverse customer workloads across analytics, search, and content understanding.
\end{abstract}



\maketitle

\section{Introduction}

The growth of large language models (LLMs) \cite{brown2020language, ouyang2022training, touvron2023llama, wei2022chain} has transformed how organizations interact with data. Modern enterprises now store vast volumes of unstructured content such as documents, images, audio, and text alongside traditional structured data in relational tables. While contemporary data warehouses excel at processing structured data with SQL, they lack native support for semantic reasoning over unstructured information. As a result, users who need to filter documents by topic relevance, join tables based on semantic similarity, or summarize customer feedback must export data to external systems, write custom scripts, or build complex pipelines that orchestrate LLM APIs outside the database. The friction is significant: data must be moved across systems, increasing latency and cost; manual orchestration becomes error-prone; and the database loses the ability to optimize the workload end-to-end.

We present Snowflake's Cortex AISQL, a SQL engine that enables users to \emph{talk} to their data, both structured and unstructured, through native AI operators integrated directly into SQL. AISQL extends SQL with primitive operators that bring LLM capabilities to the relational model: \texttt{AI\_COMPLETE} for text generation, \texttt{AI\_FILTER} for semantic filtering, \texttt{AI\_JOIN} for semantic joins, \texttt{AI\_CLASSIFY} for categorization, and \texttt{AI\_AGG} and \texttt{AI\_SUMMARIZE\_AGG} for semantic aggregations and text summarizations, respectively. These operators naturally compose with traditional SQL constructs, allowing users to write declarative queries that blend relational operations with semantic reasoning. For example, a product manager analyzing customer feedback can filter support transcripts for conversations where customers expressed frustration (\texttt{AI\_FILTER}), semantically join them with a product catalog to identify which specific products were discussed (\texttt{AI\_JOIN}), classify the severity of each issue (\texttt{AI\_CLASSIFY}), and generate executive summaries grouped by product category (\texttt{AI\_SUMMARIZE\_AGG}), all within a single SQL query that combines structured sales data with unstructured conversational text.

AISQL builds on an emerging body of work in semantic query processing. Systems such as LOTUS \cite{patel2025semantic} introduced semantic operators for processing pandas-like dataframes, Palimpzest \cite{liu2025palimpzest} provided a declarative framework for LLM-powered data transformations, while ThalamusDB \cite{jo2025thalamusdb} explored multimodal query processing with natural language predicates. Commercial systems have also adopted this direction: Google BigQuery \cite{fernandes2015bigquery} provides AI functions such as \texttt{AI.CLASSIFY} for semantic operations, while Azure SQL exposes row-level LLM invocations within SQL queries \cite{azure2025aifunctions}. However, making semantic operators efficient at production scale in a distributed database remains a fundamental challenge. Such operators are orders of magnitude more expensive than traditional SQL operations. For example, a single \texttt{AI\_FILTER} predicate applied to a million-row table could invoke an LLM once per row, leading to prohibitive execution costs and latency. Traditional query optimization heuristics, such as minimizing join costs or pushing filters below joins, can produce extremely inefficient plans when AI operators are involved. Semantic operators also behave as black boxes to the query optimizer. Unlike conventional predicates, their cost and selectivity cannot be inferred from historical statistics or column distributions. Estimating how many rows will pass an AI-based filter, or how expensive that evaluation will be, is nearly impossible at compile time. Yet, these placement and cost estimates can alter overall query performance dramatically. 

The AISQL query execution engine addresses these challenges through novel optimization techniques and runtime adaptations shaped by production deployment data from Snowflake customers. LLM inference costs far exceed traditional query processing expenses, with AI operations accounting for the dominant share of execution budgets. Multi-table queries are also prevalent, representing nearly 40\% of all AISQL workloads and consuming over half of aggregate computation time (see Section~\ref{sec:workload_analysis}). These observations guide our optimization strategies applied at both the query and individual operator level:

\begin{enumerate}
    \item \textbf{AI-aware query optimization.} The AISQL optimizer treats AI inference cost as a first-class optimization objective. Rather than optimizing solely for traditional metrics like join cardinality, the optimizer considers the monetary and computational cost of LLM invocations when determining operator placement and predicate evaluation order. For instance, the optimizer may pull expensive multi-modal \texttt{AI\_FILTER} predicates above joins or reorder multiple AI predicates based on their relative cost, even when traditional selectivity-based heuristics would suggest otherwise. Across synthetic datasets and queries modeled after actual Snowflake customer workloads, we have measured 2$\times$–8$\times$ faster execution plans. During execution, runtime statistics about predicate cost and selectivity enable dynamic reordering to adapt to actual data distributions.

    \item \textbf{Adaptive model cascades.} To reduce inference costs and maintain query prediction accuracy, AISQL implements model cascades for \texttt{AI\_FILTER} operations. A fast and inexpensive proxy model processes most rows, while a powerful oracle model handles only uncertain cases. The system learns confidence thresholds online through importance sampling and adaptive threshold estimation, partitioning rows into accept, reject, and uncertainty regions~\cite{liskowski2026streamingmodelcascadessemantic}. Our algorithm is tailored for a distributed execution environment and achieves 2--6$\times$ speedups across various benchmarks while maintaining 90--95\% of oracle model quality. On the NQ dataset, the cascade achieves 5.85$\times$ speedup and preserves a near-identical F1 score; see Section~\ref{sec:experiments-model-cascades} for details.

    \item \textbf{Query rewriting for semantic joins.} Naive semantic joins using \texttt{AI\_FILTER} typically require a quadratic number of inference calls, making them impractical for large datasets. The AISQL engine rewrites certain join patterns into multi-label classification problems using \texttt{AI\_CLASSIFY} and reduces complexity from quadratic to linear. Our implementation not only achieves 15--70$\times$ speedups across a number of datasets but often improves prediction quality because the multi-label formulation enables better comparative reasoning. On the CNN dataset, the rewrite achieves 69.52$\times$ speedup, reducing execution time from 4.4 hours to 3.8 minutes; see Section~\ref{sec:experiments-join-rewrite} for details.

\end{enumerate}

AISQL is deployed in production at Snowflake, fully integrated into the SQL engine, where it powers diverse customer workloads across analytics, search, and content understanding.


\section{Architecture}\label{sec:aisql-architecture}

\begin{figure}[t]
  \flushright
        \centering
        \includegraphics[width=0.48\textwidth]{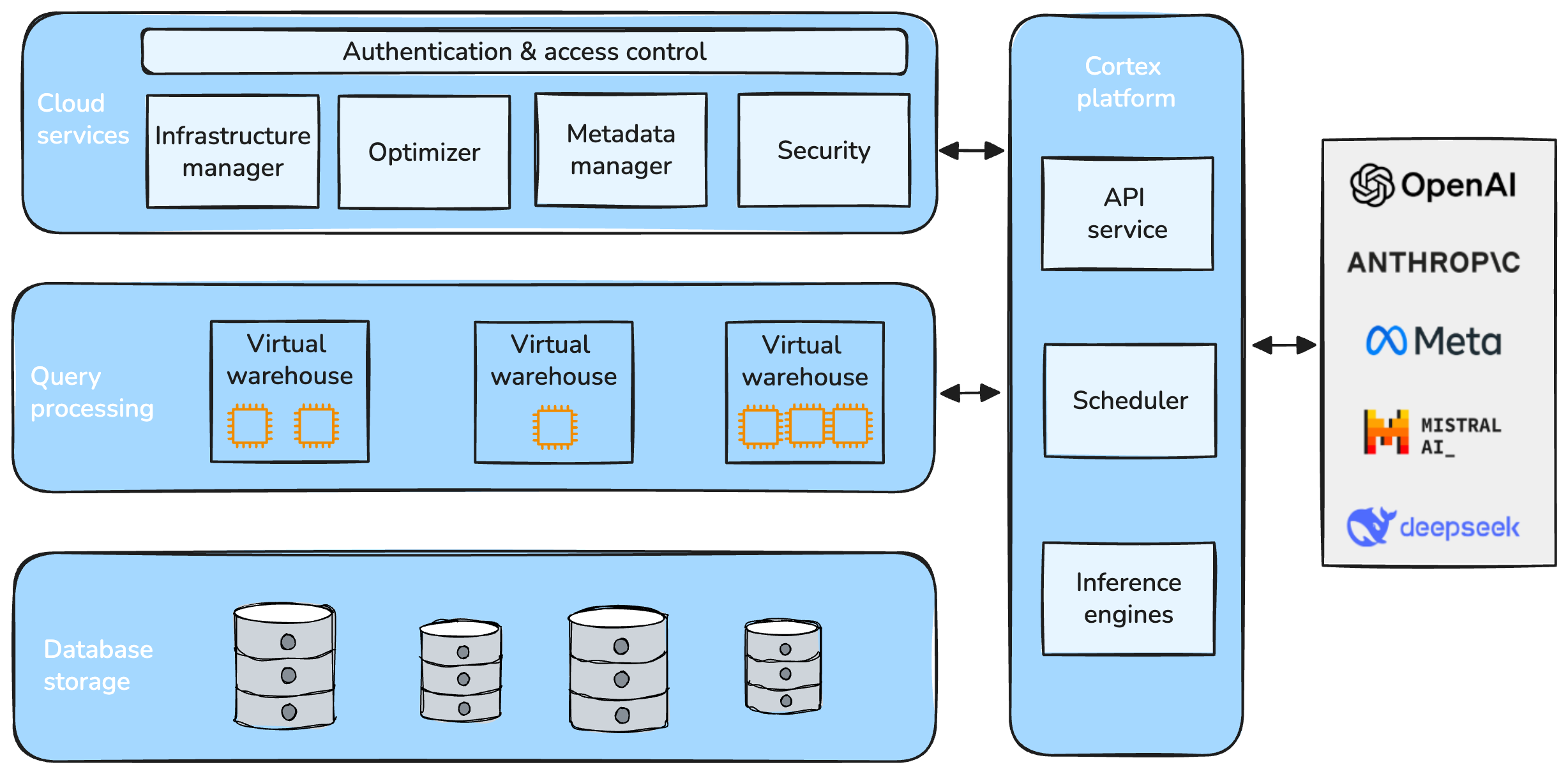}
        \caption{Snowflake architecture with the Cortex Platform for AI inference. The Cortex Platform adds Inference Engines, Scheduler, and API Service components to support both interactive and batch AISQL workloads.}
        \label{fig:ai_architecture}
\end{figure}

Snowflake's architecture has evolved to support AI workloads. Separation of compute and storage is a key design decision in Snowflake’s architecture that lets users scale query-processing resources (i.e., virtual warehouses) independently of the amount of data stored in Snowflake tables. The \textit{Cloud Services} layer handles authentication, query compilation, and coordination of SQL execution.
\begin{table*}[t]
    
    \caption{AISQL function signatures with text columns $X,Y$, predicate $\varphi$, and instruction $\varrho$.}
    \label{tab:relational_algebra}
    \centering
    \resizebox{\textwidth}{!}{
    \begin{tabular}{@{}llp{8cm}c@{}}
        \toprule
        \textbf{Operator} & \textbf{Function Signature} & \textbf{Description} & \textbf{Multimodal} \\ 
        \midrule
        $\texttt{AI\_COMPLETE}_\varrho$ 
        & $(x_1, \dots, x_n) \mapsto (\varrho(x_1), \dots, \varrho(x_n)), \quad \varrho: x_i \to y_i$ 
        & Project each row to text based on task instruction $\varrho(x)$ 
        & \checkmark \\[0.5em]
        
        $\texttt{AI\_FILTER}_\varphi$ 
        & $(x_1, \dots, x_n) \mapsto (\varphi(x_1), \dots, \varphi(x_n)), \quad \varphi: x \to \{0,1\}$ 
        & Project each row to boolean based on predicate $\varphi(x)$ 
        & \checkmark \\[0.5em]
        
        $\texttt{AI\_JOIN}_\varphi$ 
        & $(x_1, \dots, x_m), (y_1, \dots, y_n) \mapsto \{ (i, j) \mid \varphi(x_i,y_j) \}$ 
        & Select indices from $(X,Y)$ that satisfy $\varphi(x_i,y_j)$ 
        & \checkmark \\[0.5em]
        
        $\texttt{AI\_CLASSIFY}_\varrho$ 
        & \parbox[t]{7cm}{$(x_1, \dots, x_n), [c_1, \dots, c_k] \mapsto (\varrho(x_1), \dots, \varrho(x_n)),$ \\ $\quad \varrho: x_i \to c_j, \ c_j \in \{c_1, \dots, c_k\}$}
        & Project each row to discrete category $c_j$ from candidate set $\{c_1, \dots, c_k\}$ per instruction $\varrho(x)$ 
        & \\[1.5em]
        
        $\texttt{AI\_AGG}_\varrho$ 
        & $(x_1, \dots, x_n) \mapsto y, \quad \varrho: X_{i:j} \to y_k$ 
        & Reduce a column to text aggregate based on the instruction $\varrho(X)$ 
        & \\[0.5em]
        
        $\texttt{AI\_SUMMARIZE\_AGG}$ 
        & $(x_1, \dots, x_n) \mapsto y$ 
        & Reduce a column to a text summary 
        & \\[0.5em]
        \bottomrule
    \end{tabular}}
\end{table*}
The commoditization of AI, driven by the rapid growth and adoption of LLMs, has introduced new workloads and reshaped existing data-management tasks. AI agents today need to operate on data traditionally stored and managed in a warehouse. At the same time, users want to run analytical workloads and extract insights from unstructured data (e.g., documents, images, audio, and video). Section~\ref{sec:workload_analysis} analyzes AI workloads at Snowflake.

To support these workloads, we introduced the \textit{Cortex Platform} into the Snowflake architecture (Figure~\ref{fig:ai_architecture}). The Cortex Platform is a multi-tenant service that executes both interactive (e.g., via REST) and batch AI workloads (e.g., via SQL). Its primary components are (a) \textit{Inference Engines}, (b) the \textit{Scheduler}, and (c) the \textit{API Service}.

Each Inference Engine is a specialized service that hosts open-weight models (e.g., Llama, Mistral) on Snowflake-managed GPU infrastructure to generate predictions or insights from data. An engine manages both the underlying hardware and the inference stack (e.g., vLLM), distinct from the partner endpoints described below. The Cortex Platform automatically scales engines up or down to match fluctuations in inference demand. 

The Scheduler is the component responsible for orchestrating requests and assigning them to the most appropriate Inference Engine. For instance, an inference request for a specific LLM (e.g., Llama 3.1 70B) is routed to an engine that already hosts that model and is ready to serve it. 

The API Service acts as the front-end for the Cortex Platform. It receives inference requests from either the Cloud Services layer or the Query Processing layer, applies API-specific business logic, and forwards the request to the Scheduler. 

The Cortex Platform supports a broad set of models from partners including OpenAI, Anthropic, and Meta. For each inference request, the Cortex Platform determines whether to execute it using one of its Inference Engines or forward it to a partner endpoint. Requests for GPT models, for example, are routed to OpenAI's endpoints. 

\section{AISQL Operators}\label{sec:operators}

The core operators that power semantic data processing in Cortex AISQL extend SQL with natural-language semantics, allowing users to express complex analytical tasks directly within queries. For each operator, we specify its signature, behavior, and typical usage, and show how the operators compose with standard SQL while exposing data-parallel execution to support rich analytical workflows. Table~\ref{tab:relational_algebra} shows the definitions of the key semantic operators implemented in AISQL.

\subsection{Map}
AISQL supports simple map or projection operations using the $\texttt{AI\_COMPLETE}_\varrho$ operator. Users can use $\texttt{AI\_COMPLETE}_\varrho$ to transform text expressions using a low-level interface to an LLM and receive a text response for each row provided based on the task instruction $\varrho$. A $\texttt{PROMPT}$ object provides a convenience interface for including image or other multimodal data in the operator call. 

For example, users may query the system as follows: 

\begin{small}
\begin{lstlisting}[language=SQL, breaklines=true, columns=flexible, showstringspaces=false]
SELECT AI_COMPLETE(PROMPT('Evaluate the customer satisfaction from the product review: {0}', review))
FROM product_reviews;
\end{lstlisting}
\end{small}

Example result:

\begin{table}[H]
\centering
\small
\begin{tabular}{|p{0.8\columnwidth}|}
    \hline
    \textbf{AI\_COMPLETE} \\ \hline
    This review indicates moderate dissatisfaction. \\ \hline
    This review expresses positive sentiment. \\ \hline
\end{tabular}
\end{table}


\subsection{Filter}
$\texttt{AI\_FILTER}_\varphi$ is a boolean operator for use in a SQL $\texttt{WHERE}$ clause. For example, users can filter the rows of a table using a natural language predicate $\varphi$ as follows:

\begin{small}
\begin{lstlisting}[language=SQL, breaklines=true, columns=flexible, showstringspaces=false]
SELECT transcript_id, sales_agent_id 
  FROM sales_transcripts
 WHERE AI_FILTER(PROMPT('In this sales transcript,
          does the customer become irritated? {0}', 
        transcript));
\end{lstlisting}
\end{small}

\subsection{Join}
Using the $\texttt{PROMPT}$ object, the $\texttt{AI\_FILTER}_\varphi$ can be extended to multiple table and column arguments for use in a SQL $\texttt{JOIN}$. For example, users can join multiple tables using natural language as follows:

\begin{small}
\begin{lstlisting}[language=SQL, breaklines=true, columns=flexible, showstringspaces=false]
SELECT p.id, COUNT(*)
  FROM transcripts AS t
  JOIN products AS p
    ON AI_FILTER(PROMPT('In this sales transcript, does the customer complain about {0}? {1}', p.name, 
    t.transcript));
\end{lstlisting}
\end{small}

A naive implementation applies $\mathcal{O}(MN)$ \texttt{AI\_FILTER} invocations joining tables with $M$ and $N$ rows respectively, though optimization techniques help reduce this growth rate in practice (see Section~\ref{sec:ai_join}).

\subsection{Classify}
The $\texttt{AI\_CLASSIFY}_\varrho$ operator projects each row into a discrete category selected from a finite candidate set based on a natural language instruction $\varrho$. In contrast to $\texttt{AI\_COMPLETE}_\varrho$, which freely generates text, $\texttt{AI\_CLASSIFY}_\varrho$ constrains the output to one of the predefined categories $\{c_1, \dots, c_k\}$, effectively performing a supervised classification in natural language.

Similarly to $\texttt{AI\_COMPLETE}_\varrho$, each classification is computed independently, supporting distributed execution. The output column may also be used as an ordinary categorical attribute in downstream SQL operations such as \texttt{GROUP BY}, \texttt{FILTER}, or \texttt{AGG} clauses. For example, users may compute aggregate metrics per sentiment category as follows:

\begin{small}
\begin{lstlisting}[language=SQL, breaklines=true, columns=flexible, showstringspaces=false]
SELECT AI_CLASSIFY(review,['positive','neutral','negative'],
           'Classify the sentiment of this product review.') AS sentiment,
        COUNT(*) AS review_count
  FROM product_reviews
 GROUP BY sentiment;
\end{lstlisting}
\end{small}

The operator generalizes beyond sentiment analysis to any classification problem expressible in natural language (intent detection, topic tagging, severity rating, or customer status identification) while maintaining compatibility with SQL's composable semantics.

\subsection{Reduce}

\begin{algorithm}
\caption{AI\_SUMMARIZE\_AGG}
\label{alg:text_summarize}
\small
\begin{algorithmic}[1]
\Require Column of text values $texts$
\Ensure Aggregated text string
\State $R \gets \emptyset$ \Comment{Buffer for text rows}
\State $S \gets \emptyset$ \Comment{Buffer for intermediate states}

\For{$t$ in $texts$}
    \If{$(R \cup \{t\}).size() > \textsc{BATCH\_SIZE}$}
        \State $S \gets S \cup \texttt{LLM.Extract}(R)$
        \State $R \gets \emptyset$
    \EndIf
    \State $R \gets R \cup \{t\}$
    \While{$S.size() > \textsc{BATCH\_SIZE}$}
        \State $S \gets$ $\texttt{LLM.Combine}(S)$
    \EndWhile
\EndFor

\If{$R.size() > 0$}
    \State $S \gets S \cup \texttt{LLM.Extract}(R)$
\EndIf
\While{$len(S) > 1$}
    \State $S \gets$ $\texttt{LLM.Combine}(S)$
\EndWhile

\State \Return $\texttt{LLM.Summarize}(S[0])$

\end{algorithmic}
\end{algorithm}

AISQL supports two aggregate functions, $\texttt{AI\_SUMMARIZE\_AGG}$ and $\texttt{AI\_AGG}_\varrho$ that reduce a column of text values to a single aggregate result. For example, users can summarize the values of a string column as follows:

\begin{small}
\begin{lstlisting}[language=SQL, breaklines=true, columns=flexible, showstringspaces=false]
SELECT product_id,
        AI_SUMMARIZE_AGG(review)
  FROM ad_feedback
 GROUP BY product_id;
\end{lstlisting}
\end{small}

Unlike row-level operators, aggregation presents a unique challenge: tables and columns often contain more text than can fit in a single model's context window. To address this, we employ a hierarchical aggregation strategy that processes data in batches and recursively combines intermediate results.

Our approach relies on three distinct phases:
\begin{enumerate}
  \item \textbf{Extract} key information from a subset of the text rows into intermediate states
  \item Recursively \textbf{combine} intermediate states, discarding extraneous information while synthesizing similar data
  \item \textbf{Summarize} the combined state in a style and tone appropriate for the user's context
\end{enumerate}

The pseudo-code in Algorithm~\ref{alg:text_summarize} shows how the aggregate function iterates over rows, accumulating text rows in a row buffer $R$. Once the row buffer has exceed the $\textsc{BATCH\_SIZE}$ token limit, $\texttt{LLM.Extract}(R)$ generates intermediate states containing the most important information for summarization, which are all inserted into an intermediate state buffer $S$. 

Once the state buffer has exceeded $\textsc{BATCH\_SIZE}$, $\texttt{LLM.Combine}(S)$ synthesizes as many intermediate states as possible (based on the context window limit) into combined intermediate states. This process repeats until $S$ is sufficiently small.

Finally, after all rows are processed, $\texttt{LLM.Extract}$ and \texttt{LLM.Combine} are executed until $R.empty()$ and $len(S) == 1$. The final user-facing summary is generated by an LLM invocation to $\texttt{LLM.Summarize}(S[0])$.

\subsubsection{Task Specific Aggregation}

Algorithmically, $\texttt{AI\_AGG}_\varrho$ is nearly identical to $\texttt{AI\_SUMMARIZE\_AGG}$, except that each LLM request to $\texttt{LLM.Extract}$, $\texttt{LLM.Combine}$, and $\texttt{LLM.Summarize}$ supports an additional argument for a user-provided natural language task instruction $\varrho$. 

The user can provide a more specific task instruction for the aggregation as follows:

\begin{small}
\begin{lstlisting}[language=SQL, breaklines=true, columns=flexible, showstringspaces=false]
SELECT product_id, 
        AI_AGG(review, 'Identify the three most common complaints and provide recommendations to improve customer satisfaction.')
  FROM user_reviews
 GROUP BY product_id;
\end{lstlisting}
\end{small}

\subsection{Supporting Multimodal Input}
To support multimodal input such as images, audio, or documents, AISQL added a new data type, termed \texttt{FILE}. A \texttt{FILE} value stores a URI as well as various metadata (e.g. size, mime type, creation date, etc) of a file that lives in cloud storage and is managed by the user. 

Several utility functions are provided that allow users to manage and process \texttt{FILE}s. For instance, processing only the image files that are stored in a Snowflake table can be performed as follows using the boolean \texttt{FL\_IS\_IMAGE} function that checks if a file is of a known image file type:

\begin{figure}[t]
        \centering
        \includegraphics[width=0.47\textwidth]{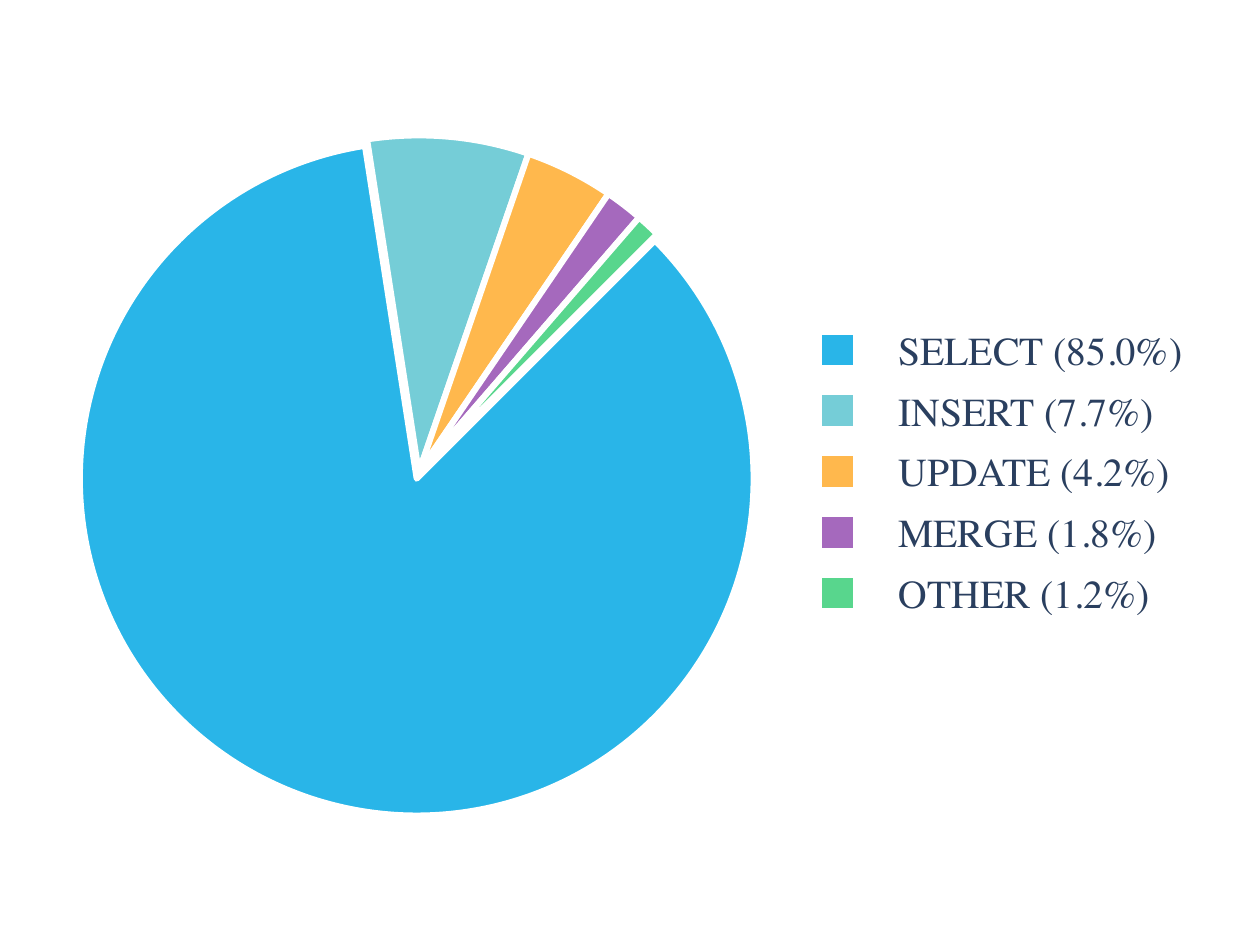}
        \caption{Percentage composition of AISQL workloads by statement type. SELECT queries constitute the majority of production workloads.}
        \label{fig:volume_by_statement}
\end{figure}

\begin{figure}[t!]
        \centering
        \includegraphics[width=0.47\textwidth]{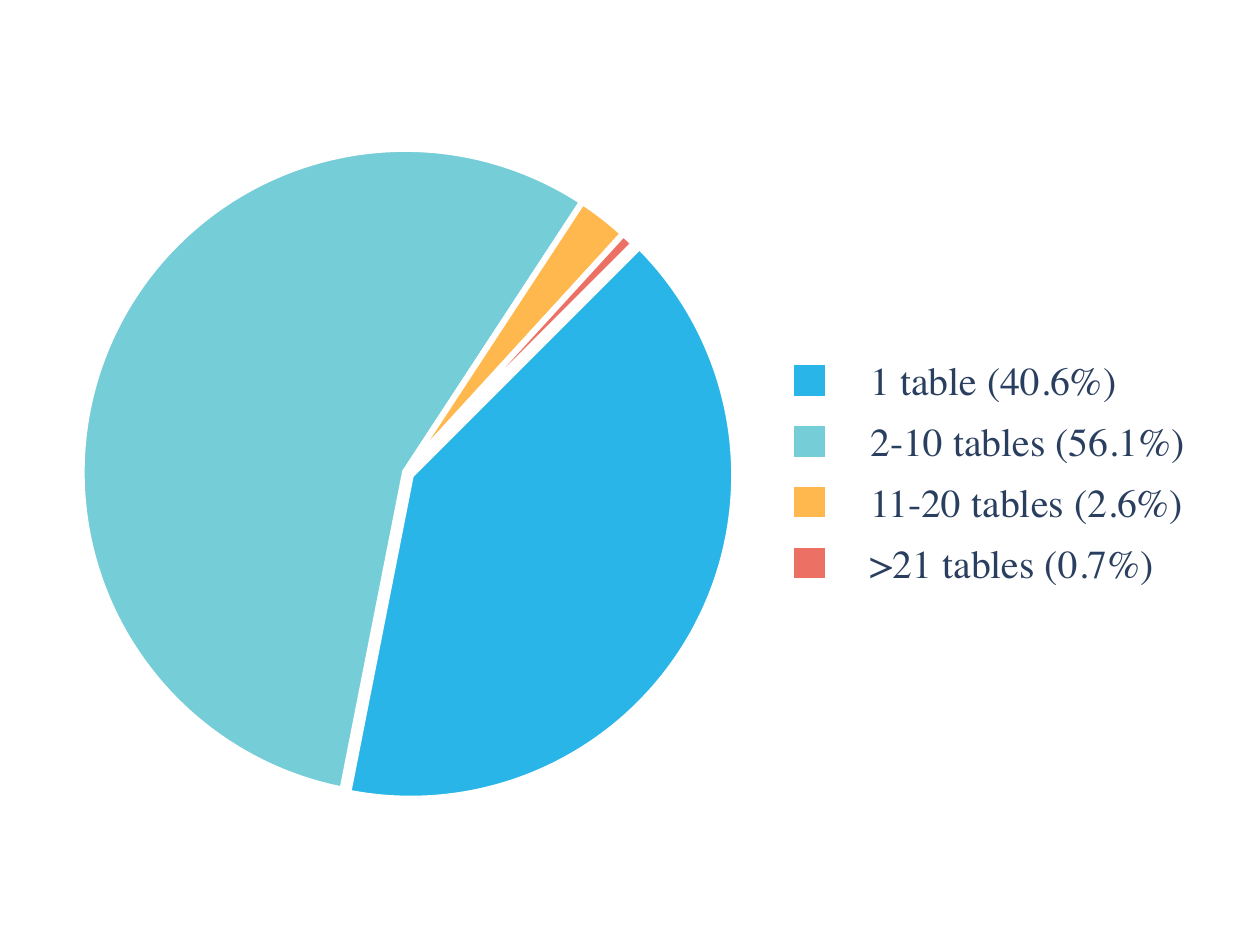}
        \caption{AISQL query execution time distribution by number of tables. Multi-table queries exhibit higher execution times than single-table queries.}
        \label{fig:exec_time_by_tables}
\end{figure}

\begin{small}
\begin{lstlisting}[language=SQL, breaklines=true, columns=flexible, showstringspaces=false]
SELECT AI_COMPLETE('claude-3-5-sonnet',
            'Identify the kitchen appliance brands from the image', 
            marketing_content.file_ref)
  FROM marketing_content
 WHERE FL_IS_IMAGE(marketing_content.file_ref);
\end{lstlisting}
\end{small}

\section{Customer Workloads}\label{sec:workload_analysis}

\begin{figure}[t]
        \centering
        \includegraphics[width=0.47\textwidth]{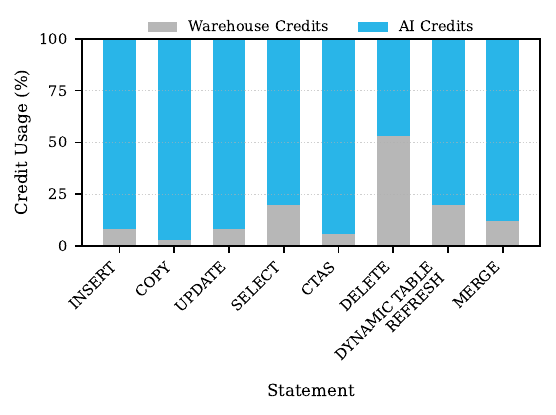}
        \caption{Cost breakdown of AISQL queries. Percentage of total credit usage by statement type, showing the relative contributions of model inference (AI credits) and relational processing (Warehouse credits).}
        \label{fig:cost_breakdown}
\end{figure}

\begin{figure}[t]
        \centering
        \includegraphics[width=0.45\textwidth]{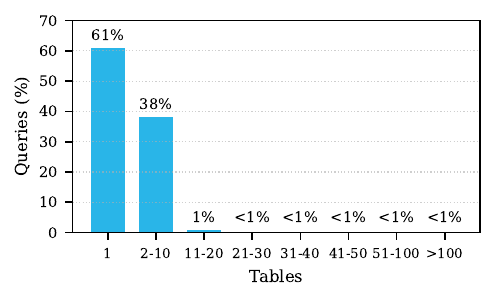}
        \caption{Distribution of tables used in AISQL queries. Most AISQL queries ($61\%$) involve a single table, while multi-table queries account for nearly $40\%$ of workloads.}
        \label{fig:perc_by_tables}
\end{figure}

As discussed earlier, executing semantic operations inside a database engine introduces several system-design challenges. At the same time, customers are adapting to the AI-centric data management era and exploring new workloads and usage patterns to extract maximum value from AI. We analyze AISQL workloads executed over a three-month period (July–September 2025) across multiple Snowflake deployments and cloud providers. The observations below are intended to inform practitioners and motivate further research. Our main observations from the workload analysis are as follows:

\begin{enumerate}
    \item {\bf AI operators dominate AISQL query cost.} The observation follows from the analysis of AISQL query composition by statement type (see Figure~\ref{fig:volume_by_statement}) and the cost distribution between LLM GPU credits and SQL Warehouse credits for each statement type (see Figure~\ref{fig:cost_breakdown}). Thus, in the next section, we present optimization techniques that reduce the cost of running AISQL queries with 
    (a) {\it AI-Aware Query Optimization} that takes into account the high cost of executing AI operators to create an effective query plan (see Section~\ref{sec:optimizer}); and 
    (b) {\it Adaptive Model Cascades} that uses a lightweight proxy LLM to process most rows while using a larger oracle LLM to handle difficult cases where the proxy model is uncertain (see Section~\ref{sec:ai_filter}). 

    \item {\bf Semantic joins are prevalent in customer workloads.}
    \newline Nearly $40\%$ of AISQL queries involve multiple tables (see Figure~\ref{fig:perc_by_tables}), and these multi-table queries account for over $58\%$ of total execution time (see Figure~\ref{fig:exec_time_by_tables}). Thus, we propose a new method for optimizing semantic joins that reduces the cost and latency of the operation (see Section~\ref{sec:ai_join}).
\end{enumerate}

\section{AISQL Query Execution Engine}\label{sec:ai_sql_engine}

Snowflake's core database engine required several modifications to support AISQL queries, i.e. SQL queries that use one or more AI operators. Section~\ref{sec:optimizer} presents query plan optimizations tailored to AISQL workloads. Section~\ref{sec:ai_filter} discusses the use of model cascade techniques to improve the efficiency of the \texttt{AI\_FILTER} operator. Section~\ref{sec:ai_join} details an optimization that rewrites semantic joins into a multi-label classification problem using \texttt{AI\_CLASSIFY}, and Section~\ref{sec:ai_agg} covers enhancements applied to aggregation operators such as \texttt{AI\_AGG} and \texttt{AI\_SUMMARIZE\_AGG}. Predicate reordering and cost-based placement adapt well-known optimization techniques for expensive predicates~\cite{hellerstein6caching, hellestein93predicate, chaudhuri99optimization} to the AISQL setting. The remaining two techniques are novel contributions: adaptive model cascades introduce a streaming threshold-learning algorithm with formal precision-recall guarantees~\cite{liskowski2026streamingmodelcascadessemantic}, and the join-to-classification rewrite exploits the structure of semantic join predicates to reduce quadratic-cost joins to linear classification.

\subsection{Optimizing AI Operators}\label{sec:optimizer}
One of the key challenges in optimizing AISQL queries is that AI operators are treated as black boxes by the optimizer. For example, without historical data from previous invocations on the same data, it is not possible to estimate the selectivity and hardware cost of an \texttt{AI\_FILTER} predicate. At the same time, the per-row monetary and runtime costs of AI operators are orders of magnitude higher than those of conventional SQL operators and functions. Consequently, the suboptimal placement of an AI operator in the execution plan can have a dramatic impact on both cost and execution time. The problem is not new in database systems. Optimizing user-defined functions (UDFs) in database engines shares many similarities with the problem of optimizing AI operators, and many of the techniques published 20 years ago remain relevant \cite{hellerstein6caching, hellestein93predicate, chaudhuri99optimization}.

\begin{figure}[thbp]
        \centering
        \includegraphics[width=0.48\textwidth]{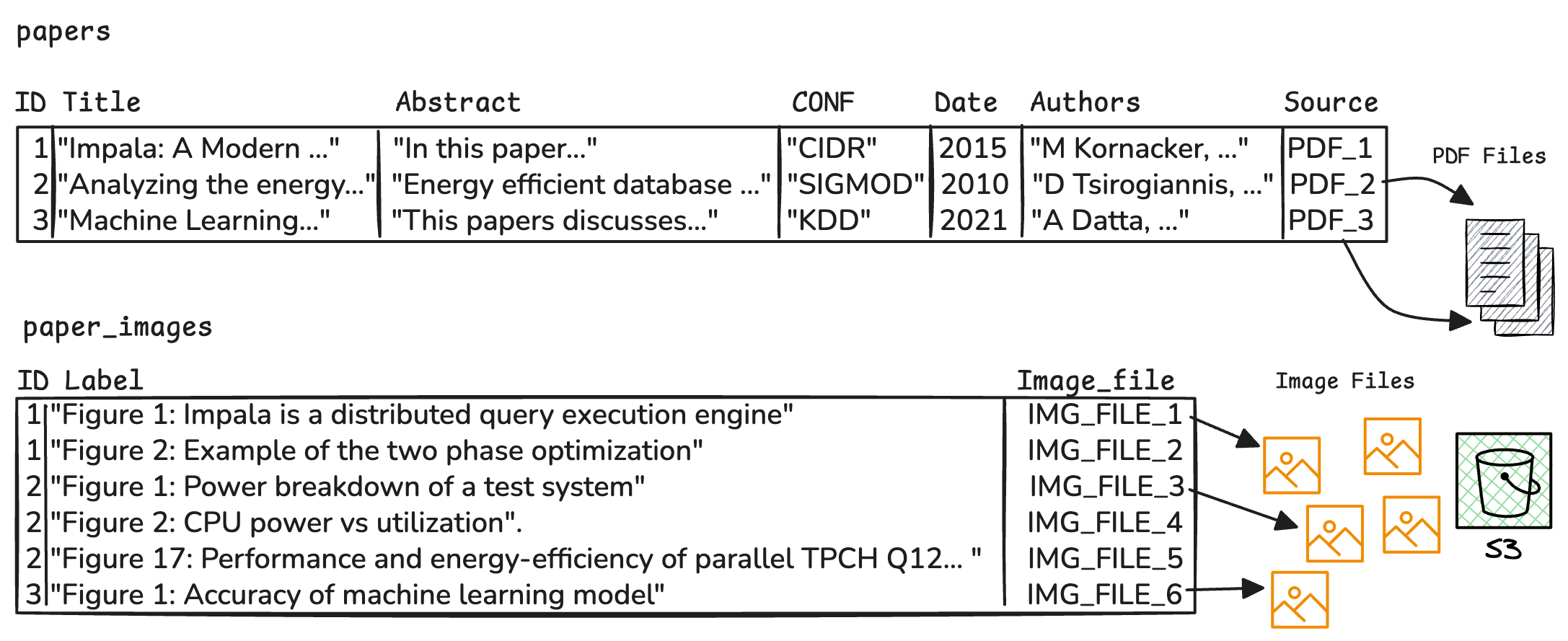}
        \caption{Schema of research papers application.}
        \label{fig:qp_example}
\end{figure}

To illustrate the challenges of optimizing AISQL queries, let us consider the following scenario: a web application like \textit{arxiv.org} manages research papers and allows users to search for papers using semantic operations. For simplicity, we assume that the content of each research paper is stored in two relational tables \textit{papers} and \textit{paper\_images}. The former stores the basic information for each paper, such as "title" and "authors", whereas the latter stores information about extracted images and graphs for each paper. Both tables have a column of type \textit{FILE} that links each row with a file stored in cloud storage (e.g., AWS S3). A simplified schema and some sample data are shown in Figure~\ref{fig:qp_example}.

Now let us consider a user who is interested in identifying the research papers that: a) were published in a specific time period, b) discuss "energy efficiency in database systems", and c) include a figure with results using the TPC-H benchmark. Finally, the user is interested in compiling a summary of the corresponding abstracts. This query requires inspecting both the textual and image contents of research papers. Today, the user can write the following AISQL query in Snowflake:

\smallskip
\begin{minipage}{\linewidth}
\begin{small}
\begin{lstlisting}[language=SQL, caption={Example AISQL query with two AI functions, AI\_FILTER and AI\_SUMMARIZE\_AGG}.]
SELECT AI_SUMMARIZE_AGG(p.abstract) 
FROM papers p JOIN paper_images i ON p.id = i.id 
WHERE p.date between 2010 and 2015 AND 
    AI_FILTER(PROMPT('Abstract {0} discusses energy efficiency in database systems', p.abstract)) 
    AND AI_FILTER(PROMPT('Image {0} shows energy consumption of different systems using the TPC-H workload', 
    i.image_file));
\end{lstlisting}
\end{small}
\end{minipage}

The AISQL query in this example has three predicates with different cost and performance characteristics. The first is a predicate on a "date" column, which is relatively "cheap" to evaluate and for which most database engines can reasonably estimate its selectivity. The second predicate is an \texttt{AI\_FILTER} on a VARCHAR column, which is more expensive as it invokes an LLM for each row and its selectivity is unknown at compile time. The execution cost can be estimated based on the average number of tokens in the column values. The third predicate is also an \texttt{AI\_FILTER}, but it is applied to images and invokes a multimodal model. Multimodal models are generally larger and more expensive than models operating only on text input.

\begin{figure}[thbp]
        \centering
        \includegraphics[width=0.48\textwidth]{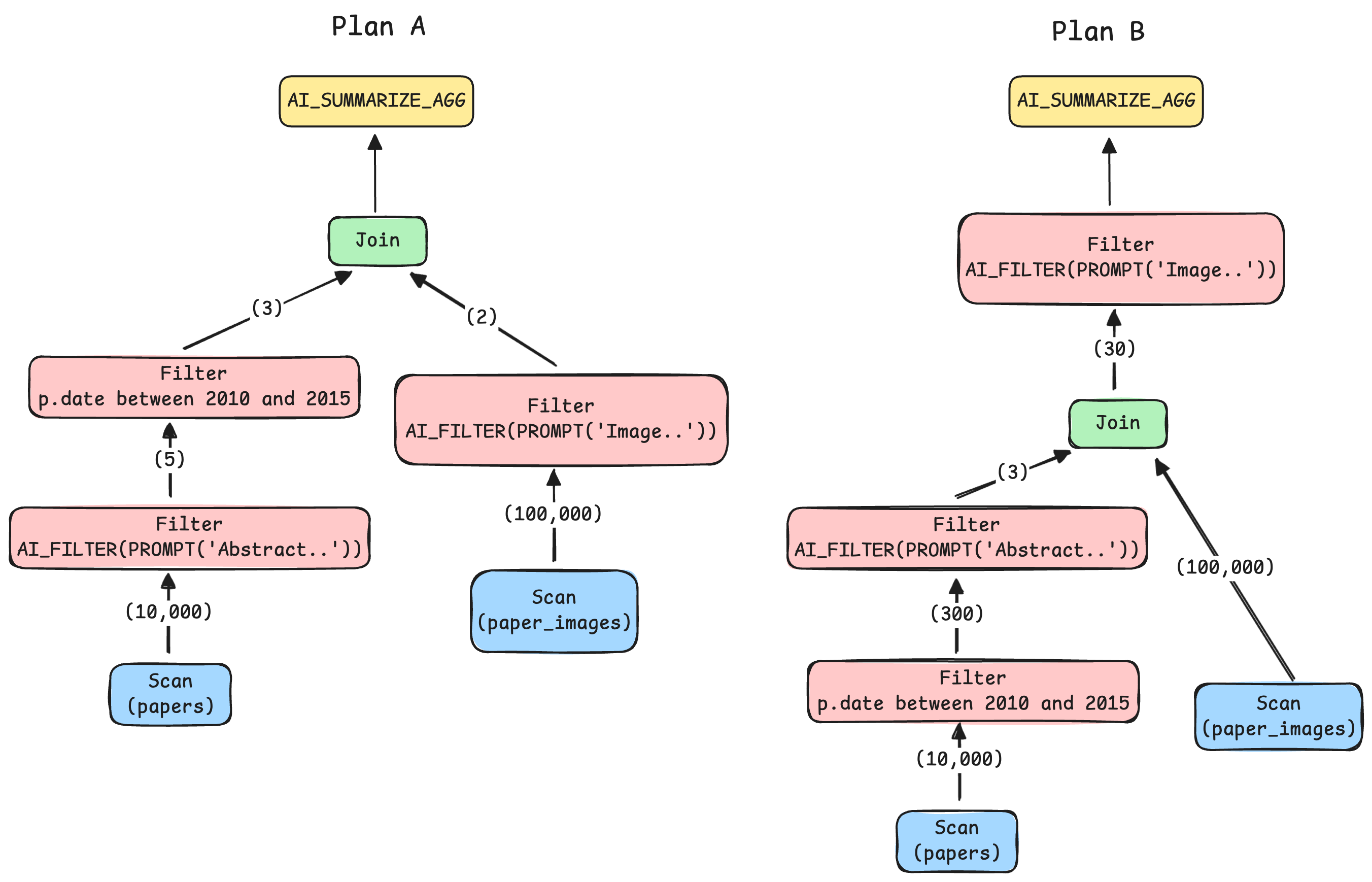}
        \caption{Different execution plans. Plan A optimizes for join cost. 
        Plan B considers the cost of AI operators and optimizes for LLM cost.}
        \label{fig:qp_execution_plan}
\end{figure}

Suppose the cost of each predicate is ignored. In that case, the optimizer may generate \textit{Plan A}, shown in Figure~\ref{fig:qp_execution_plan}, in which all predicates are "pushed" below the join, with the underlying assumption that join is an expensive operator and hence the number of rows sent to this operator should be minimized. The most selective predicate (\texttt{AI\_FILTER} in this example) is applied first on the rows scanned from the "papers" table. \textit{Plan A} results in 110,000 LLM calls; the numbers in the parentheses indicate the output cardinality of each operator.

Figure~\ref{fig:qp_execution_plan} shows a better execution plan (\textit{Plan B}) that takes into account the cost of AI predicates and tries to minimize the number of LLM calls. By "pulling" the image \texttt{AI\_FILTER} predicate above the join and by changing the order in which predicates are applied on the rows from the "papers" table, the compiler generates an execution plan with only 330 LLM calls, a 300$\times$ improvement in both cost and execution time. 

The example highlights two of the optimizations that the Snowflake execution engine applies to improve the performance of AISQL queries (see Section~\ref{sec:experiments} for an experimental evaluation), namely: a) predicate reordering, b) optimizing query plans for AI inference cost. Predicate reordering has both a compiler and an execution component. During compilation, the predicate evaluation order is determined based on the relative cost of each predicate, i.e., the most expensive predicates are evaluated last. During execution, runtime statistics are collected on the cost and selectivity of predicates. These statistics are used to determine at runtime whether a different predicate evaluation order is more effective. For instance, if an AISQL query evaluates two \texttt{AI\_FILTER} predicates on text columns from the same input, we can determine at runtime which one is more selective and change the evaluation order so that the more selective \texttt{AI\_FILTER} predicate is applied first.
 
As we discussed in Section~\ref{sec:workload_analysis}, AISQL queries spend a significant fraction of execution time in AI operators. Consequently, optimizing AISQL queries requires us to rethink the optimization criteria used in the compiler. We initially considered how \texttt{AI\_FILTER} predicates are placed in the execution plan with respect to joins and aggregations, as this problem showed up in numerous customer workloads. Although the selectivity of an \texttt{AI\_FILTER} is unknown at compile time, we identified that simply optimizing the total number of AI inference calls produces good results in many cases. However, as AISQL queries become more complex and compiler estimations less reliable, such optimizations become less effective in practice. To that end, we are considering dynamic optimization techniques, caching runtime statistics, and methods to generalize AI function placement in complex execution plans. 

\subsection{Adaptive Model Cascades}\label{sec:ai_filter}

AI operators such as $\texttt{AI\_FILTER}$ must often process millions of rows in production workloads. Invoking an LLM on every row introduces prohibitive costs and latency. The AISQL engine addresses this with \emph{model cascades}: a lightweight \emph{proxy model} (e.g., Llama3.1-8B) processes all rows and produces a confidence score for each, while a powerful \emph{oracle model} (e.g., Llama3.3-70B) handles only cases where the proxy is uncertain.

\smallskip
\noindent\textbf{Confidence Scores.} For each row $x_i$, the proxy model produces a score $s_i \in [0,1]$ representing the estimated probability that the predicate is satisfied. The score is derived from the model's output logits: for an $\texttt{AI\_FILTER}$ predicate, the proxy generates a binary (yes/no) response and $s_i$ is the softmax probability assigned to the positive-class token.

\smallskip
\noindent\textbf{Two-Threshold Routing.} Two learned thresholds $\tau_{\text{low}}$ and $\tau_{\text{high}}$ partition rows into three regions based on their proxy scores:
\begin{enumerate}
    \item \textbf{Reject} ($s_i < \tau_{\text{low}}$): predict negative without oracle evaluation.
    \item \textbf{Accept} ($s_i \geq \tau_{\text{high}}$): predict positive without oracle evaluation.
    \item \textbf{Uncertainty} ($\tau_{\text{low}} \leq s_i < \tau_{\text{high}}$): route to the oracle for reliable classification.
\end{enumerate}
By classifying high- and low-confidence rows with the proxy alone, the cascade reserves oracle invocations for the uncertainty region and the importance sample used for threshold learning.

\smallskip
\noindent\textbf{Adaptive Threshold Learning.} AISQL deploys the SUPG-IT algorithm~\cite{liskowski2026streamingmodelcascadessemantic}, which extends the SUPG statistical framework~\cite{kang2020approximate} to streaming execution with joint precision-recall guarantees. Data is partitioned across parallel workers, each processing its batches independently without inter-worker communication. Within each batch, the algorithm samples a budget fraction $\rho$ of rows from the full batch for oracle labeling using importance sampling with weights proportional to $\sqrt{s_i}$, combined with uniform mixing for coverage. The accumulated oracle labels drive iterative threshold refinement: $\tau_{\text{low}}$ is set from a weighted ROC curve with a sampling-corrected recall target, while $\tau_{\text{high}}$ is the minimum threshold whose statistical lower bound on precision meets the precision target. As oracle samples accumulate across batches, confidence bounds tighten, the uncertainty region narrows, and fewer rows require oracle evaluation. Rows remaining in the uncertainty region are routed to the oracle if budget permits; otherwise, the proxy prediction serves as a fallback.

In production, the cascade operates transparently: users issue standard \texttt{AI\_FILTER} queries and the engine automatically activates proxy-oracle routing. Users can optionally specify an oracle budget fraction and precision or recall targets to control the cost-quality tradeoff. After each query, the system reports the observed delegation rate so users can adjust these parameters iteratively. Section~\ref{sec:experiments-model-cascades} evaluates the cascade's speed-quality tradeoff across six benchmarks in Snowflake's production environment. A detailed treatment of SUPG-IT and a complementary calibration-based cascade algorithm, including formal guarantees and convergence analysis, is presented in~\cite{liskowski2026streamingmodelcascadessemantic}.

\subsection{Query Rewriting for Semantic Joins}\label{sec:ai_join}

\begin{figure}[thbp]
        \centering
        \includegraphics[width=0.48\textwidth]{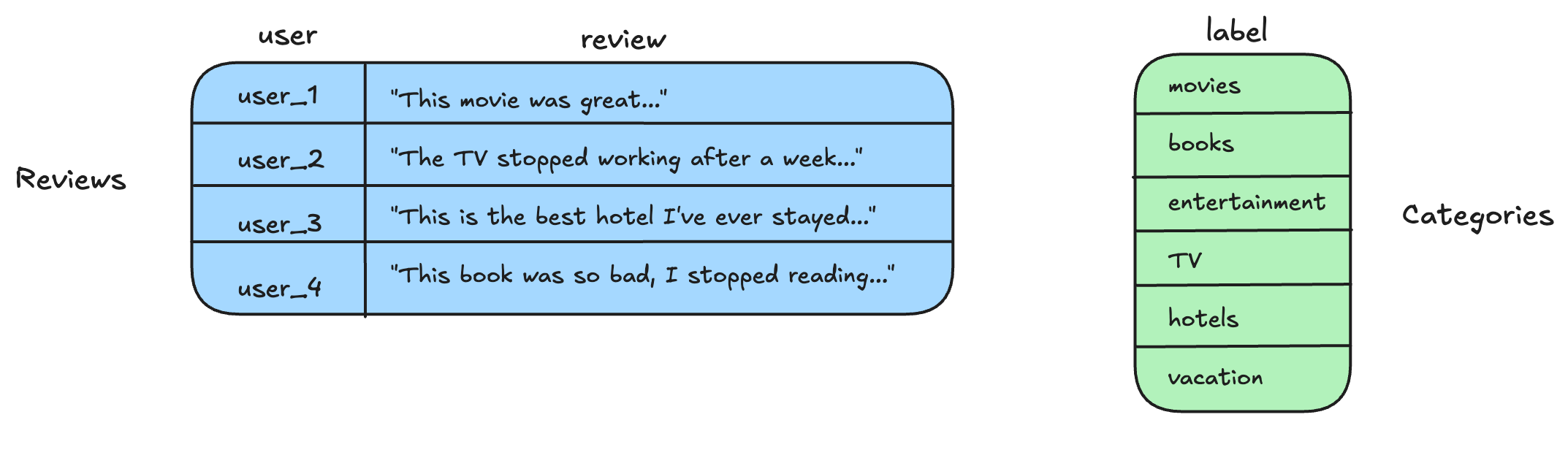}
        \caption{Example of two Snowflake tables with sample data.}
        \label{fig:ai_join_example}
\end{figure}

Semantic joins using \texttt{AI\_FILTER} enable natural language-based table joining. For example, let us consider a scenario where a user has two Snowflake tables: (1) “Reviews” contains product reviews, and (2) “Categories” includes names of product categories. A simplified schema of these two tables, along with some sample data, is shown in Figure~\ref{fig:ai_join_example}.

Consider an example where a user wants to associate every product review with one or more categories. The user can write the following AISQL queries that join \textit{Reviews} with \textit{Categories} using an \texttt{AI\_FILTER} predicate in the ON clause.

\smallskip
\begin{minipage}{\linewidth}
\begin{small}
\begin{lstlisting}[language=SQL, caption={Example AISQL query that performs a semantic join between two tables.}]
SELECT * FROM
Reviews JOIN Categories
ON AI_FILTER(PROMPT('Review {0} is mapped to category {1}',
    Reviews.review, Categories.label));
\end{lstlisting}
\end{small}
\end{minipage}

Due to the semantic nature of the join condition, efficient join algorithms such as hash-join or sort-merge join, which rely on equality predicates, are not applicable. The AISQL execution engine must therefore treat the semantic join as a cross join, thereby generating all possible pairs from \textit{Reviews} and \textit{Categories} and evaluating \texttt{AI\_FILTER} for each pair. Therefore, such queries can produce thousands of LLM calls and high inference costs, even for small datasets. More specifically, the above approach requires $O(|L| \times |R|)$ LLM calls, where $|L|$ and $|R|$ denote the cardinalities of the left and right tables, respectively. 

After analyzing a wide range of customer queries, we discovered that, in many cases, semantic joins are equivalent to a multi-label classification problem. For instance, in the previous example, rather than applying \texttt{AI\_FILTER} for every review-category pair, we can use the Snowflake \texttt{AI\_CLASSIFY} function, which can classify a text or image input into one or more user-specified labels. By re-framing the semantic join as a classification problem, we can instead execute \texttt{AI\_CLASSIFY} for every row of table \textit{Reviews} using the values of column \textit{Categories.label} as labels, thereby reducing the number of AI inference calls from 24 (4 reviews $\times$ 6 labels) to just 4 (one for each review), a 6$\times$ improvement in the example above.
 
Automatically detecting such cases is not a trivial task. During the compilation of AISQL queries, we introduced an AI-based oracle that examines every semantic join to determine if it can be rewritten as a multi-label classification problem. The oracle analyzes the user-specified natural language prompt, schema metadata (e.g., column and table names), statistics (e.g., number of distinct values), as well as sample values from each input source. Using all that information, it determines if a semantic join should be converted into a classification task and which input dataset contains the labels for the classification. Subsequently, the transformation is performed as a regular rewrite operation by the compiler. The rewritten query is equivalent to the original user query but adds additional operations to: a) reduce the number of input labels to each \texttt{AI\_CLASSIFY} call, and b) handle cases where the input of each \texttt{AI\_CLASSIFY} call fits in the context window of the user-specified LLM model. Section~\ref{sec:experiments-join-rewrite} evaluates the rewrite on eight benchmarks.

\subsection{AI Aggregation}\label{sec:ai_agg}

Earlier, we described an incremental fold strategy for combining intermediate states within a Reduce operation for both $\texttt{AI\_AGG}_\varrho$ and $\texttt{AI\_SUMMARIZE\_AGG}$. However, invoking three separate LLM APIs for $\texttt{LLM.Extract}$, $\texttt{LLM.Combine}$, and $\texttt{LLM.Summarize}$ introduces significant overhead for small datasets that fit within the model context window. Therefore, a simple "short-circuit" alternative improves throughput by identifying these scenarios and skipping unnecessary incremental fold steps. The optimization achieved an 86.1\% latency reduction in $\texttt{AI\_SUMMARIZE\_AGG}$ on queries using small datasets.

\section{Experimental Evaluation}\label{sec:experiments}
All experiments run on a production-release version of Snowflake. We evaluate three optimization techniques on public benchmarks: AI-aware query optimization (Section~\ref{sec:experiments-query-optimization}), adaptive model cascades (Section~\ref{sec:experiments-model-cascades}), and query rewriting for semantic joins (Section~\ref{sec:experiments-join-rewrite}). For each technique, we measure execution time and prediction quality against baseline approaches.

\subsection{AI-aware Query Optimization}\label{sec:experiments-query-optimization}
\begin{figure}[t]
  \includegraphics[width=0.48\textwidth]{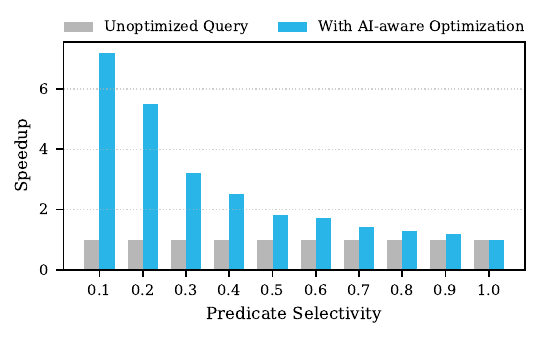}
  \caption{Effect of predicate reordering on the performance of AISQL queries that include both AI and non-AI predicates.}
  \label{fig:predicate_reordering}
\end{figure}

In this section, we evaluate the performance impact of two compiler optimizations: a) predicate reordering and b) LLM cost-optimized predicate placement with respect to joins. Figure~\ref{fig:predicate_reordering} shows the performance impact of predicate reordering for an AISQL query shown below. The SQL statement has two predicates in the WHERE clause, an \texttt{IN} predicate and an \texttt{AI\_FILTER} predicate on a VARCHAR column. We used a dataset with 1000 articles from the New York Times\footnote{https://www.nytimes.com/}. 

\begin{minipage}{\linewidth}
\begin{small}
\begin{lstlisting}[language=SQL, ]
SELECT year, title FROM NYT_ARTICLES
  WHERE id_group IN (<list of group ids>) AND AI_FILTER(PROMPT('The article title is about finance: {0}', title), {'model': 'llama3.1-70b'});
\end{lstlisting}
\end{small}
\end{minipage}

In this experiment, we vary the selectivity of the \texttt{IN} predicate between $0.1$ and $1$, with $1$ indicating that all input rows satisfy the \texttt{IN} predicate. We then measure the speedup from reordering the predicate evaluation so that the \texttt{AI\_FILTER} predicate is evaluated last. The normalized results in Figure~\ref{fig:predicate_reordering} show that applying the \texttt{AI\_FILTER} last yields up to a 7$\times$ speedup in our experiment. In practice, the actual speedup will depend on several factors, such as the relative costs of the different predicates and their selectivities. 

In the second experiment, we measure the impact on total execution time of the compiler optimization that decides the placement of AI predicates (\texttt{AI\_FILTER}) with respect to joins based on the total LLM cost. We consider the following AISQL statement that performs a join between two input tables and has two predicates in the ON clause, one of which is an \texttt{AI\_FILTER}.

\begin{small}
\begin{lstlisting}[language=SQL, ]
SELECT * FROM NYT_ARTICLES_V1 AS l JOIN NYT_ARTICLES_V2 AS r ON l.id = r.id AND 
AI_FILTER(PROMPT('The article title is about finance: {0}', l.title), {'model': 'llama3.1-70b'});
\end{lstlisting}
\end{small}

For this experiment, we adjust the ratio of input rows from the left input (NY\_ARTICLES\_V1), where the $\texttt{AI\_FILTER}$ is applied, to the total number of output rows generated by the join. We vary the ratio between $0.1$ and $2$, where $2$ means that the join produces twice as many rows as the cardinality of the left input. We compare the optimization that considers the total LLM cost when deciding the placement of \texttt{AI\_FILTER} against two other approaches: a) \textit{Always Pull-up} and, b) \textit{Always Push-down}. \textit{Always Pull-up} always "pulls" AI predicates on top of joins, whereas \textit{Always Push-down} is the default behavior in Snowflake's optimizer that always "pushes" predicates below joins. 

\begin{figure}[t]
  \includegraphics[width=0.48\textwidth]{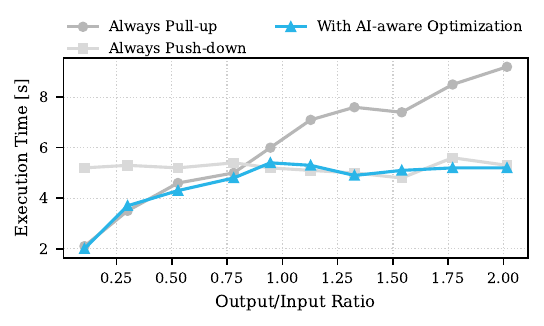}
  \caption{Effect of optimizing AI predicates with respect to joins.}
  \label{fig:ai_filter_and_joins}
\end{figure}

\begin{figure*}[t]
  \centering
  \includegraphics[width=\textwidth]{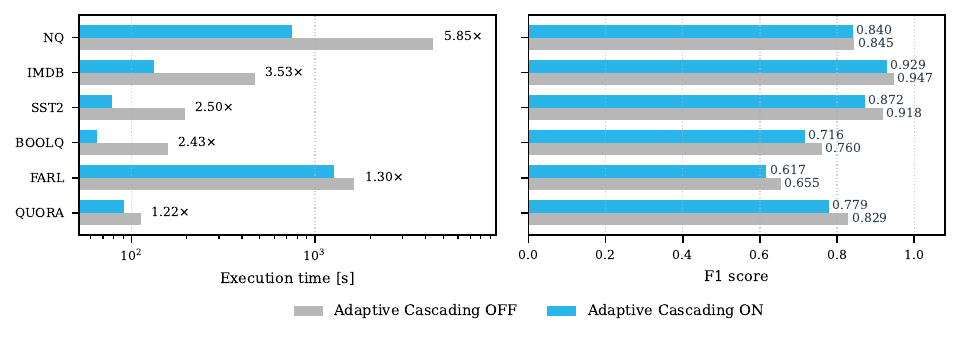}
  \caption{Performance comparison of adaptive model cascades on six benchmark datasets. The cascade uses adaptive threshold learning to route predictions between a proxy model (Llama3.1-8B) and oracle model (Llama3.3-70B). \textbf{Left:} Execution time (in seconds) with speedup factors annotated. \textbf{Right:} F1 scores with exact values labeled. The cascade reduces execution time by 1.2--5.9$\times$ with average F1 decreasing by 4.3\%.}
  \label{fig:model_cascades_performance}
\end{figure*}

Figure~\ref{fig:ai_filter_and_joins} shows that neither \textit{Always Pull-up} nor \textit{Always Push-down} is optimal across the different output/input ratios. When the ratio is $<1$ (i.e., a selective join), \textit{Always Pull-up} performs better. In contrast, \textit{Always Push-down} is better when the output/input ratio is $>1$ (i.e., explosive joins). Considering the total cost of LLM when deciding the placement of \texttt{AI\_FILTER} predicates (termed \textit{AI-aware Optimization}) achieves the best performance across the entire range of measured output/input ratios. In practice, the effectiveness of this technique relies on the compiler's ability to accurately estimate join selectivity, which is not a trivial problem, especially in complex queries with many joins. Future efforts will explore adaptive optimization techniques, which involve evaluating various execution plans during compilation while collecting runtime statistics, such as the cost and selectivity of AI operators, from sample data to improve decisions regarding AI operator placement. 

\subsection{Adaptive Model Cascades}\label{sec:experiments-model-cascades} 

We evaluate the performance impact of model cascades in \texttt{AI\_FILTER} operations across six Boolean classification datasets. The six datasets are public NLP benchmarks available on HuggingFace, spanning question answering (NQ, BOOLQ), sentiment analysis (IMDB, SST2), duplicate question detection (QUORA), and news veracity classification (FARL). Dataset sizes range from approximately 3,500 to 400,000 rows. As detailed in Section~\ref{sec:ai_filter}, the cascading approach uses adaptive threshold learning to route predictions between a lightweight proxy model and a powerful oracle model. We compare three configurations: (1) the baseline approach using only the oracle model Llama3.3-70B for all predictions, (2) the cascade approach combining both models with adaptive threshold learning, and (3) using only the proxy model Llama3.1-8B for all predictions. 

We execute each query five times and report the mean. We measure both execution efficiency (query time) and prediction quality using standard classification metrics: accuracy, precision, recall, and F1 score. All quality assessments use ground-truth labels from the evaluation datasets as a reference. 

Table~\ref{tab:ai_filter_cascading} reports the mean performance across all datasets, including execution time, speedup relative to the baseline, F1 score with delta ($\Delta$) relative to the baseline, and precision/recall metrics. All speedup percentages and delta F1 values are computed with respect to the baseline approach.

\begin{table}[ht]
    \centering
    \caption{Mean performance comparison of \texttt{AI\_FILTER} with different model configurations.}
    \label{tab:ai_filter_cascading}
    \resizebox{\columnwidth}{!}{
    \begin{tabular}{lrrrrr}
    \toprule
    \multirow{2}{*}{\textbf{Method}} & \multirow{2}{*}{\textbf{Time [s]}} & \multirow{2}{*}{\textbf{Speedup}} & \multicolumn{2}{c}{\textbf{F1}} & \multirow{2}{*}{\textbf{Prec. / Rec.}} \\
    \cmidrule(lr){4-5}
    & & & \textbf{Score} & $\mathbf{\Delta}$ & \\
    \midrule
    llama3.3-70B & 975.9 & --- & 0.812 & --- & 0.813 / 0.829 \\
    llama3.1-8B & 296.2 & 3.3x & 0.659 & -18.8\% & 0.704 / 0.686 \\
    Cascade  & 336.4 & 2.9x & 0.777 & -4.3\% & 0.784 / 0.794 \\
    \bottomrule
    \end{tabular}
    }
    \end{table}

\noindent\textbf{Performance gains.} The cascade approach achieves substantial speedup with mean execution time reduced by 65.5\% (from 975.9 to 336.4 seconds), corresponding to a 2.9$\times$ speedup. The improvement stems from routing the majority of predictions through the lightweight proxy model, reserving the oracle model only for uncertain cases. For comparison, the proxy model alone runs in 296.2 seconds (3.3$\times$ speedup), as expected from its smaller model size, though at a significant quality cost.

Figure~\ref{fig:model_cascades_performance} shows per-dataset results, revealing that speedup varies considerably across workloads. The NQ dataset achieves the highest speedup at 5.85$\times$, while QUORA and FARL show moderate improvements (1.22$\times$ and 1.30$\times$ respectively). These variations reflect differences in how the threshold learning algorithm partitions the data. Easier workloads, where the proxy model exhibits stronger confidence scores, yield higher routing rates to the proxy and thus greater speedups.

\noindent\textbf{Quality analysis.} The cascade maintains strong prediction quality with only modest mean degradation. The F1 score decreases by 4.3\% (from 0.812 to 0.777), with balanced precision and recall remaining at 0.784 and 0.794, respectively. The result confirms that the threshold learning logic identifies rows where the proxy model's confidence scores are reliable. In contrast, using the proxy model exclusively degrades quality considerably, with F1 dropping by 18.8\% to 0.659, underscoring that the oracle model is essential for handling predictions in the uncertainty region.

As shown in Figure~\ref{fig:model_cascades_performance}, the impact on quality scores varies across datasets. NQ maintains near-identical F1 ($0.845 \rightarrow 0.840$), while BOOLQ experiences the largest drop ($0.760 \rightarrow 0.716$). IMDB, SST2, and QUORA fall between these extremes. The variations suggest that cascade effectiveness depends on dataset characteristics: tasks where the proxy model's uncertainty correlates well with prediction difficulty benefit most from the adaptive routing strategy.

\noindent\textbf{Discussion.} The cascade offers an attractive balance between the baseline and proxy-only approaches. For applications where near-optimal quality is essential, the baseline achieves the highest F1 score (0.812) but at a 2.9$\times$ higher computational cost. For latency-critical workloads where moderate quality is acceptable, the cascade delivers 65.5\% speedup while retaining 95.7\% of baseline F1 performance. The proxy-only configuration is suitable only when speed is paramount and substantial quality degradation is tolerable.



\subsection{Query Rewriting for Semantic Joins}\label{sec:experiments-join-rewrite}

\begin{figure*}[!ht]
  \centering
  \includegraphics[width=\textwidth]{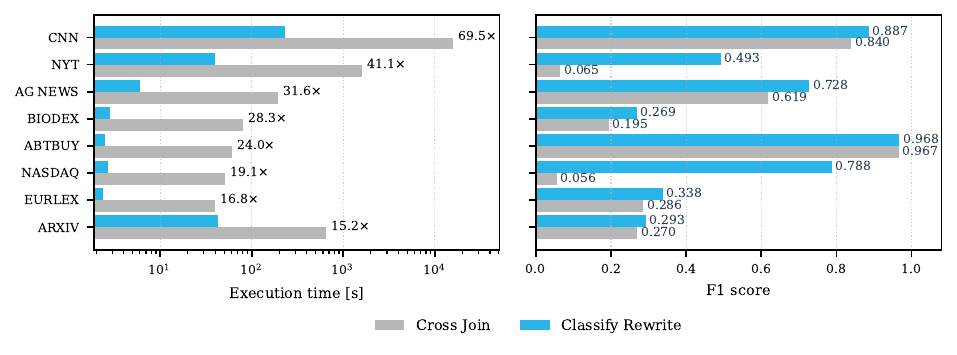}
  \caption{Performance comparison of \texttt{AI\_CLASSIFY} rewrite optimization on eight semantic join benchmarks. \textbf{Left:} Execution time (in seconds) with speedup factors annotated. \textbf{Right:} F1 scores with exact values labeled. The rewrite reduces execution time by 15.2--69.5$\times$ with mean F1 improving by 44.7\%.}
  \label{fig:rewrite_performance}
\end{figure*}
We evaluate \texttt{AI\_CLASSIFY} rewrite optimization using eight semantic join benchmarks, all publicly available on HuggingFace, that cover entity matching (ABTBUY, NASDAQ), document categorization (ARXIV, EURLEX, NYT, CNN), news classification (AG~NEWS), and biomedical concept linking (BIODEX). Each benchmark consists of two tables with natural language content, where the join task is to identify semantically related row pairs. For example, in ABTBUY the predicate matches product descriptions from different e-commerce sites that refer to the same item, while in NYT it links news articles to their topical categories. The left table (denoted as $L$ in Table~\ref{tab:ai_join_rewrite}) and the right table ($R$) span 50 to 500 rows, yielding up to 250,000 candidate pairs that must be considered by the naive join approach using filtering. Table~\ref{tab:ai_join_rewrite_mean} summarizes the mean performance across all benchmarks. 
\begin{table}[ht]
    \centering
    \caption{Mean performance comparison between the Cross Join with \texttt{AI\_FILTER} and the \texttt{AI\_CLASSIFY} rewrite across all datasets.}
    \label{tab:ai_join_rewrite_mean}
    \resizebox{\columnwidth}{!}{
    \begin{tabular}{lrrrrr}
    \toprule
    \multirow{2}{*}{\textbf{Method}} & \multirow{2}{*}{\textbf{Time [s]}} & \multirow{2}{*}{\textbf{Speedup}} & \multicolumn{2}{c}{\textbf{F1}} & \multirow{2}{*}{\textbf{Prec. / Rec.}} \\
    \cmidrule(lr){4-5}
    & & & \textbf{Score} & $\mathbf{\Delta}$ & \\
    \midrule
    Cross Join & 2330.56 & --- & 0.412 & --- & 0.388 / 0.761 \\
    Classify Rewrite & 40.96 & 30.7x & 0.596 & +44.7\% & 0.745 / 0.540 \\
    \bottomrule
    \end{tabular}
    }
    \end{table}

\noindent\textbf{Performance gains.} The query rewriting achieves large speedups across all benchmarks, ranging from $15.24\times$ (ARXIV) to $69.52\times$ (CNN), with a mean speedup of $30.7\times$. These improvements stem directly from reducing LLM invocations: the baseline requires $O(|L| \times |R|)$ calls, while the rewrite needs only $O(|L|)$ classifications. For the CNN dataset with 500 rows per table, this reduces 250,000 binary classifications to just 1,000 multi-label classifications, cutting execution time from 4.4 hours to 3.8 minutes. The speedup magnitude correlates strongly with dataset scale; larger Cartesian products yield proportionally greater benefits.

\begin{table*}[ht]
\centering
\small
\caption{Performance comparison between the Cross Join with \texttt{AI\_FILTER} and the \texttt{AI\_CLASSIFY} rewrite across all datasets.}
\label{tab:ai_join_rewrite}
\begin{tabular}{l r r c@{\hspace{3pt}}c@{\hspace{3pt}}c@{\hspace{3pt}}c@{\hspace{3pt}}c c@{\hspace{3pt}}c@{\hspace{3pt}}c@{\hspace{3pt}}c@{\hspace{3pt}}c c}
\toprule
\multirow{2}{*}{\textbf{Dataset}} & \multirow{2}{*}{\textbf{L}} & \multirow{2}{*}{\textbf{R}} & \multicolumn{5}{c}{\textbf{Cross Join (AI\_FILTER)}} & \multicolumn{5}{c}{\textbf{AI\_CLASSIFY Rewrite}} & \multirow{2}{*}{\textbf{Speed}} \\
\cmidrule(lr){4-8} \cmidrule(lr){9-13}
 & & & \textbf{Calls} & \textbf{Time} & \textbf{Precision} & \textbf{Recall} & \textbf{F1} & \textbf{Calls} & \textbf{Time} & \textbf{Precision} & \textbf{Recall} & \textbf{F1} & \\
\midrule
NASDAQ & 100 & 100 & 10000 & 51.46 & 0.029 & 0.96 & 0.056 & 100 & 2.69 & 0.851 & 0.731 & 0.788 & $19.13\times$ \\
EURLEX & 50 & 194 & 9700 & 39.93 & 0.172 & 0.833 & 0.286 & 50 & 2.37 & 0.86 & 0.21 & 0.338 & $16.84\times$ \\
BIODEX & 50 & 197 & 9850 & 79.6 & 0.118 & 0.585 & 0.195 & 50 & 2.81 & 0.409 & 0.2 & 0.269 & $28.32\times$ \\
ABTBUY & 100 & 100 & 10000 & 60.27 & 0.967 & 0.967 & 0.967 & 100 & 2.51 & 0.968 & 0.968 & 0.968 & $24.01\times$ \\
AG NEWS & 100 & 100 & 10000 & 63.21 & 0.565 & 0.87 & 0.685 & 100 & 2.57 & 0.91 & 0.61 & 0.731 & $24.59\times$ \\
AG NEWS & 200 & 200 & 40000 & 192.27 & 0.505 & 0.8 & 0.619 & 200 & 6.08 & 0.905 & 0.61 & 0.728 & $31.62\times$ \\
ARXIV & 500 & 500 & 250000 & 646.67 & 0.55 & 0.18 & 0.27 & 1500 & 42.31 & 0.549 & 0.2 & 0.293 & $15.24\times$ \\
NYT & 500 & 500 & 250000 & 1618.58 & 0.034 & 0.775 & 0.065 & 1500 & 39.39 & 0.609 & 0.414 & 0.493 & $41.09\times$ \\
CNN & 500 & 500 & 250000 & 15955.67 & 0.729 & 0.99 & 0.84 & 1000 & 229.48 & 0.807 & 0.984 & 0.887 & $69.52\times$ \\
\bottomrule
\end{tabular}
\end{table*}

\noindent\textbf{Quality analysis.} The impact of rewrite optimization on prediction quality varies by dataset characteristics. On entity matching tasks with clear semantic signals (ABTBUY), both approaches achieve near-identical F1 scores (${\approx}0.97$), demonstrating that the rewrite preserves quality when matches are unambiguous. The rewrite also substantially improves quality on datasets where the baseline suffers from poor precision. For NASDAQ, the extremely low precision of the baseline ($0.029$) results in an F1 score of $0.056$, while the rewrite achieves a precision of $0.851$ and an F1 score of $0.788$. We attribute this to the multi-label classification paradigm: by presenting all candidate labels simultaneously, \texttt{AI\_CLASSIFY} enables better comparative reasoning than isolated binary decisions in \texttt{AI\_FILTER}. The mean F1 improves by $44.7\%$ (from $0.412$ to $0.596$), with precision increasing from $0.388$ to $0.745$.

\noindent\textbf{Trade-offs.} The rewrite exhibits precision-recall trade-offs that depend on the classification strategy. Datasets like EURLEX and BIODEX show recall degradation ($0.833\rightarrow0.21$ and $0.585\rightarrow0.2$, respectively) despite improved precision. Recall degradation occurs when the model conservatively selects matches from the full set of labels, prioritizing accuracy over coverage. For applications where recall is critical, hybrid strategies that combine both approaches or tune the classification prompt may be necessary.
\section{Related Work}\label{sec:related_work}

The AI4DB line of research applies ML algorithms to internal database components such as query optimization, indexing, and configuration tuning. Early systems like LEO \cite{stillger2001leo} used feedback learning to refine cost estimates for the DB2 optimizer. Later, OtterTune \cite{vanaken2017ottertune} and CDBTune \cite{zhang2019end} used large-scale supervised and reinforcement learning to automatically tune database parameters, inspiring the vision of self-driving databases \cite{pavlo2017self}.

A rich body of work explores learned query processing components. Learned Indexes \cite{kraska2018case} replaced traditional index structures with neural networks, while ALEX \cite{ding2020alex} extended these ideas to support dynamic workloads. In query optimization, learned cardinality estimation models, such as MSCN \cite{kipf2018learned}, Naru \cite{yang2019deep}, and DeepDB \cite{hilprecht2020deepdb}, achieved notable improvements in selectivity prediction by learning correlations across attributes and tables. Reinforcement learning has also been applied to plan search and join ordering, exemplified by ReJOIN \cite{marcus2018deep}, Neo \cite{marcus2019neo}, and Bao \cite{marcus2021bao}. These works culminate in the SageDB \cite{kraska2019sagedb}, which proposes a learned database system where each component is automatically specialized for a given data distribution and workload.

The application of ML to query optimization extends to handling expensive predicates in specialized domains. Systems such as BlazeIt \cite{kang2019blazeit} and TASTI \cite{kang2020tasti} optimize queries involving expensive ML-based predicates using model cascades, semantic indexes, and query-specific proxy models. Model cascades that combine cheap approximate models with expensive accurate ones \cite{kang2020approximate, chen2023frugal, zellinger2025rational} have influenced the design of AISQL's approach to managing expensive AI operations.


Earlier in-database ML frameworks such as MADLib \cite{hellerstein2012madlib} extended SQL with support for descriptive statistics and simple ML methods. Recent systems go further by integrating semantic and language-model capabilities directly into query processing. LOTUS \cite{patel2025semantic} introduces a set of semantic operators for Pandas-like dataframe processing. ThalamusDB \cite{jo2025thalamusdb} explores approximate query processing for multimodal data, supporting natural language predicates over visual, audio, and textual content through zero-shot models combined with relational operators. Unlike these systems, AISQL operates within a production distributed SQL engine and co-optimizes query planning with LLM inference cost.

Several systems have proposed specialized approaches to LLM-powered SQL processing. SUQL \cite{liu2024suql} augments SQL with \texttt{answer} and \texttt{summary} operators for knowledge-grounded conversational agents, with the focus on row-wise LLM operations for question answering. ZenDB \cite{lin2024towards} optimizes SQL queries for extracting structured data from semi-structured documents with predictable templates. Palimpzest \cite{liu2025palimpzest} provides a declarative framework for LLM-powered data processing with specialized map-like operations and basic cost-based optimizations. UQE \cite{dai2024uqe} studies embedding-based approximations for LLM-powered filters and stratified sampling for aggregations, though it provides best-effort performance without accuracy guarantees. AISQL goes beyond row-level processing with cross-row optimizations (cascades, join rewriting) that reduce inference calls by orders of magnitude.

Beyond SQL-centric approaches, several frameworks have emerged for LLM-powered data processing. Aryn \cite{anderson2024design} offers a Spark-like API with PDF extraction and human-in-the-loop processing capabilities. DocETL \cite{shankar2025docetl} proposes agent-driven pipeline optimization for complex document processing tasks and uses LLM agents to explore task decomposition strategies. EVAPORATE \cite{arora2023language} specializes in extracting semi-structured tables through code synthesis. These systems demonstrate the potential of LLM-powered processing, but often lack formal cost models and provide limited performance guarantees.

Optimizing expensive predicates and user-defined functions has been extensively studied~\cite{hellerstein6caching, hellestein93predicate, chaudhuri99optimization} and directly inspired AISQL's query optimizations. Prior work addressed cost-based predicate reordering~\cite{hellestein93predicate}, caching of expensive function results~\cite{hellerstein6caching}, and integrating user-defined predicates into cost-based optimization frameworks~\cite{chaudhuri99optimization}. AISQL extends these techniques to LLM-powered operators, where per-row costs are orders of magnitude higher and selectivities are unknown at compile time. 


\section{Conclusions}\label{sec:conclusions}
We have presented Snowflake AISQL, a production SQL engine that bridges the gap between structured and unstructured data processing by integrating LLM capabilities directly into SQL. AISQL introduces six semantic operators that compose naturally with relational SQL primitives, enabling users to express complex analytical tasks that blend structured queries with semantic reasoning in a single declarative statement.

The core technical challenge we address is making semantic operators efficient at production scale. Our solution comprises three techniques driven by analysis of real customer workloads: AI-aware query optimization that treats LLM inference cost as a first-class objective; adaptive model cascades that reduce costs through intelligent routing between lightweight proxy and powerful oracle models; and query rewriting for semantic joins that transforms quadratic-complexity joins into linear multi-label classification tasks with improved prediction quality.

Our production deployment at Snowflake validates these design choices, and experimental evaluation demonstrates that AISQL delivers both efficiency and quality at production scale. AISQL represents a step toward unified data processing architectures where semantic reasoning over unstructured content is a first-class database capability. Organizations can query both structured and unstructured data through familiar declarative interfaces while maintaining the performance and reliability expected from modern data warehouses.

Several directions for future work remain. Caching runtime statistics across queries would improve plan quality for recurring AISQL workloads. Extending model cascades beyond \texttt{AI\_FILTER} to multi-class operators requires generalizing the binary threshold framework to handle distinct confidence distributions per class. Hybrid join strategies that combine classification-based rewriting with filtering could improve recall on datasets where the rewrite alone sacrifices coverage.

\bibliographystyle{ACM-Reference-Format}
\balance
\bibliography{sigproc}
\end{document}